\documentclass[letter]{aa} 

\usepackage{graphicx}
\usepackage{txfonts}

\usepackage{graphicx}
\usepackage{natbib}
\usepackage{color}
\usepackage[colorlinks=true]{hyperref}
\bibpunct{(}{)}{;}{a}{}{,} 
\begin{document}

   \title{Discs are born eccentric}

   \subtitle{}

   \author{ Benoît Commerçon\inst{1}, 
          Francesco Lovascio\inst{1}, 
          Elliot Lynch
          \inst{1}, 
          \and
          Enrico Ragusa
          \inst{2,3}
          }

 \institute{Univ Lyon, Univ Lyon1, Ens de Lyon, CNRS, Centre de Recherche Astrophysique de Lyon UMR5574, F-69230, Saint-Genis,-Laval, France\
              \email{benoit.commercon@ens-lyon.fr},
         \and
             Dipartimento di Matematica, Università degli Studi di Milano, Via Saldini 50, 20133, Milano, Italy
             \and
              Dipartimento di Fisica, Università degli Studi di Milano, Via Celoria 16, 20133 Milano MI, Italy
             }

   \date{}

 
  \abstract
   {Recent observations have begun probing the early phases of disc formation, but little data yet exists on disc structure and morphology of Class 0 objects. Using simulations, we are  able to lay out predictions of disc morphologies expected in future surveys of young discs. Based on detailed simulations of ab initio star formation by core collapse, we predict that early discs must be eccentric.}
   {In this letter, we study the morphology and, in particular, the eccentricity of discs formed in non-ideal magnetohydrodynamic (MHD) collapse simulations. We attempt to show that discs formed by cloud collapse are likely to be eccentric.}
   {We ran non-ideal MHD collapse simulations in the adaptive mesh refinement code {\ttfamily RAMSES~} with radiative transfer. We used state-of-the-art analysis methods to measure the disc eccentricity.}
   {We find that despite no asymmetry  in the initial conditions, the discs formed are eccentric, with eccentricities on the order of 0.1}
   {These results may have important implications for protoplanetary disc dynamics and planet formation. The presence of eccentricity in young discs that is not seen at later stages of disc evolution is in tension with current viscous eccentricity damping models. This implies that there may be an as-yet undiscovered circularisation mechanism in circumstellar discs.}


   \maketitle
%

\section{Introduction}
Advances in observational and computational technologies have shed light on the properties of young protostellar discs over the past ten years. At the Class 0 stage in low-mass protostars, these discs are born compact and are regulated by non-ideal magnetohydrodynamic (MHD) processes \citep[see][for a review]{tsukamoto:23}. Recently, the large programme titled Early Planet Formation in Embedded Disks (eDisk), conducted with the Atacama Large Millimeter/submillimeter Array (ALMA), reported 1.3 mm continuum observations of Class 0/I discs at a resolution of $\sim 7$~au. The youngest Class 0 discs do not exhibit clear distinctive substructures \citep[rings and spirals][]{ohashi:23}, in contrast to Class II disks \citep{andrews:18}.  Yet, some of these young discs exhibit marked asymmetry along their major and minor axes \citep{zamponi:21,thieme:23,vanthoff:23}. We propose that this may be explained on the basis of disc eccentricity.
The eccentricity of discs has been studied theoretically and computationally over the past decades, recent developments in instruments, most notably ALMA \citep{andrews:18}, have made the detection of eccentricity in better resolved discs possible \citep[as has also been done for debris discs, see  e.g. ][]{MacGregor_2022}. In this letter, we show that young discs are expected to have a non-zero (potentially measurable) eccentricity from birth to at least the end of the accretion phase.  
\section{Methods}
This paper uses the \texttt{RAMSES} code \citep{teyssier:02} to simulate non ideal MHD collapse with the adaptive-mesh-refinement (AMR).  We ran simulations of dense core collapse down to AU scales to study the morphology of early protoplanetary disk evolution on kyr timescales. 
\subsection{Physics and numerical setup}
Our numerical framework in \texttt{RAMSES} integrates  the equation of radiation-MHD and is similar to the one used in   \cite{mignon:21a,mignon:21b}. We used a non-ideal MHD solver that accounts for ambipolar diffusion \citep{masson:12} and employs the constrained transport algorithm originally developed for ideal MHD by \cite{fromang:06} and \cite{teyssier:06}. The ambipolar diffusion resitivities are computed using a table from \cite{marchand:16}. We note that we do not account for Ohmic diffusion due to i) the high computational cost at high resolution in the inner disc and ii) the lower amplitude of Ohmic diffusion with respect to ambipolar diffusion at first Larson core scale.  For the radiative transfer, we used the hybrid irradiation solver developed by \cite{mignon:20} which combines the M1 method to handle stellar irradiation \citep{rosdahl:13,rosdahl:15} and the grey flux-limited-diffusion (FLD) solver \citep{commercon:11a,commercon:14} to handle photon emission after stellar irradiation absorption and regular photons (from compressional heating and friction due to ambipolar diffusion). Overall, our numerical setup is very similar to that of \cite{lee:21}, except for the stellar irradiation, the latter use the FLD to handle both stellar photons and regular thermal emission.  

\subsection{Initial conditions and refinement strategy}
Our simulations follow the isolated collapse of a top-hat mass distribution (uniform density) similar to those run in 
\citet{hennebelle:20} and \cite{lee:21}.  For our initial conditions we use a total cloud mass of 1 $\mathrm{M}_\odot$ and cloud radius of $R_0=3980$~au with periodic boundaries in a $16 000\mathrm{au}$ box. This was chosen to prevent interactions with the boundaries causing spurious effects in the box. We also added a 10\% azimuthal $m=2$ density perturbation.  The gas is initially at uniform temperature $T_0=10$~K, with a mean molecular weight of $\mu_\mathrm{gas}=2.31$.  The ratio between thermal and gravitational energies is $\alpha=0.4$. 
Our cloud is initially in solid body rotation with a ratio of rotational to gravitational energies $\beta=0.04$. The magnetic field is purely vertical (aligned with the $z-$axis) and has a strength which corresponds to a ratio of the initial mass-to-flux ratio over the critical mass-to-flux ratio of 3. We set a $10^{\circ}$ initial misalignment between the magnetic field $\mathbf{B}$ and the axis of rotation $\hat{\mathbf{L}}$.
The initial AMR level is $\ell_\mathrm{min}=6$, which corresponds to a resolution of $64^3$. The grid is refined according to the Jeans length criterion using 20 points per Jeans length up to the maximum level of refinement (see Sect.~\ref{sec:zoomin}). To resolve the disk with high resolution,  we cap the sound speed when the temperature gets higher than 100 K (see Fig.~\ref{fig:disc_dens}). 

\subsection{Zoom-in strategy\label{sec:zoomin}}
Even with adaptive mesh refinement, it is not feasible to resolve all the disc at our target resolution $\sim 0.25$~au or AMR level of 16, while running a simulation for extended time. Therefore, we took a "zoom-in" approach, where a fiducial simulation is evolved at lower resolution (i.e. spatial resolution $\Delta x \sim 1.0 $~au or AMR maximum level of 14) for extended time. From this simulation, we launched zoom-ins, where we restarted the simulation with a higher target resolution. The zoom-in shown here is launched at $t=12500\mathrm{yr}$ after sink formation and is allowed to evolve for $2.3~\mathrm{kyr}$, the low-resolution simulation is allowed to evolve a further $6~\mathrm{kyr}$ after the start point of the zoom-in.


\subsection{Sink physics and pre-main sequence evolution}
We used a sink particle to describe the evolution of the collapsing gas below the maximum resolution, as well as the radiative feedback from the accreting protostar.
The sink is created when a density threshold is reached and conditions on thermal support are met, the density threshold for sink formation used is  $n>1.5\times10^{13}\mathrm{cm}^{-3}$, and the conditions for thermal support are the same as in \citet{lee:21}. If the material is then deemed to be dense enough ($n>0.5\times10^{13}~\mathrm{cm}^{-3}$) and not thermally supported,  a fraction $c_\mathrm{acc}$=0.1  of the  material above the threshold is accreted into the sink \citep[for more information, see][note:\ we used the same sink model, with a lower accretion threshold to facilitate sink formation]{hennebelle:20}. For the zoom-in run, we used the same sink model, but to account for the higher resolution and to reduce sink accretion in general, we raised the accretion threshold to $n>5\cdot10^{15}~\mathrm{cm}^{-3}$, which quickly ends accretion onto the sink and, in fact, in the studied high-resolution snapshots, the sink accretion rate is $0~\mathrm{M}_\odot \mathrm{yr}^{-1}$.  With our zoom-in strategy, we tested the accreting and non-accreting sink.  \cite{hennebelle:20} showed that the disk properties, mainly its mass, depend heavily on the sink particle accretion scheme. \cite{kley:08} also reported  a strong dependence of the disk eccentricity and precession rate on the inner boundary condition. 
We assumed that a single protostar forms within the sink particle and that it  radiates as a blackbody. To characterise the stellar luminosity, we used the pre-main sequence tracks from \cite{kuiper:13} that gives the protostar radius and effective temperature. Our model for PMS evolution is identical as the one used in \cite{mignon:21a,mignon:21b}.

\subsection{Measuring the eccentricity} \label{sec:methods:ecc}
\begin{figure}
    \centering
    \includegraphics[scale=0.1]{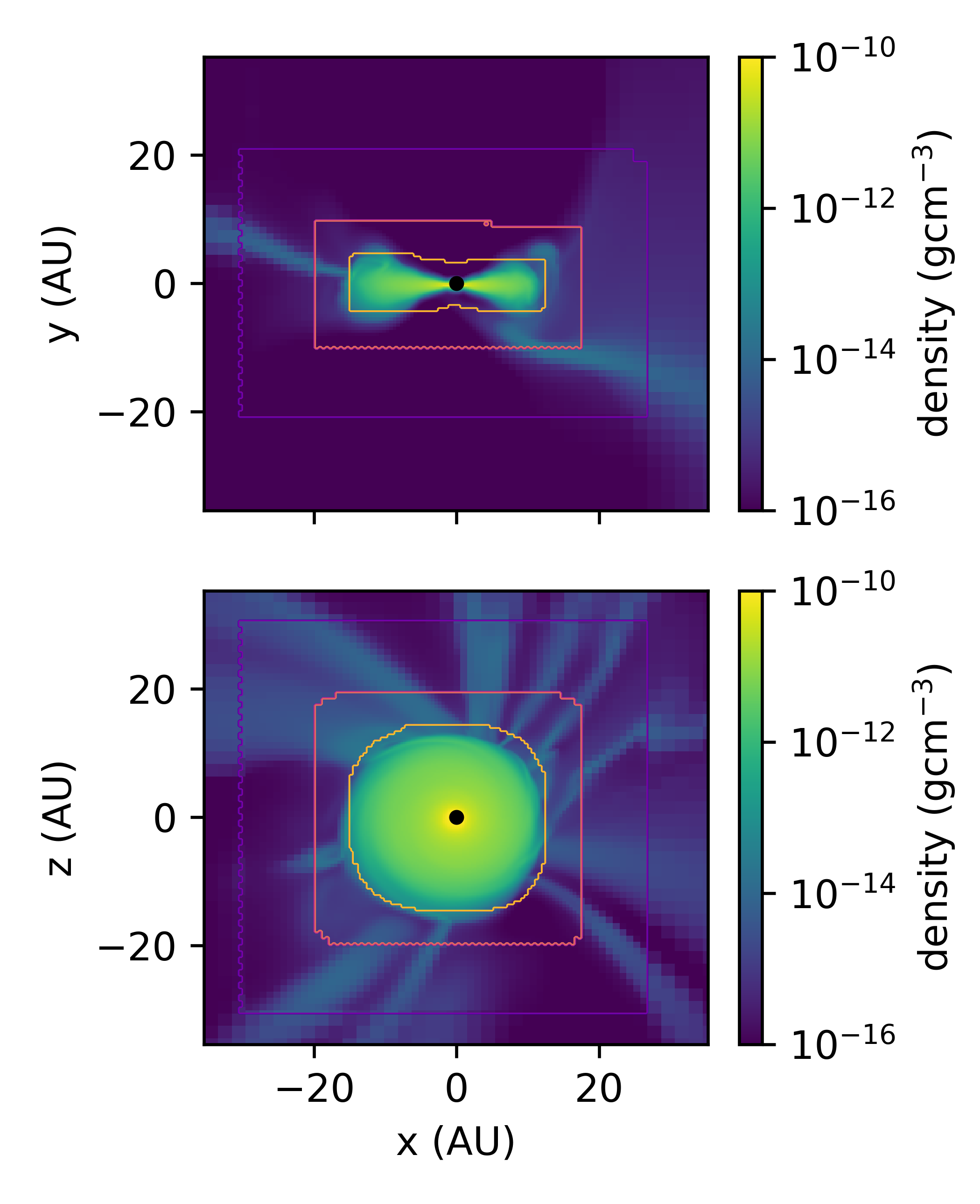}
    \caption{Density distribution 15 kiloyear after sink formation and 2 kiloyear after zoom-in, obtained from zoom-in simulation (0.25~au resolution in the disc). The disc eccentric structure is apparent from both the ellipticity of the disc in the bottom (face-on) panel, as well as the scale height variation in the top (side-on) panel. The black dot marks the sink. The contours indicates the AMR level, with the disc sitting at the highest refinement level.} 
    \label{fig:disc_dens}
\end{figure}

This study aims to determine the eccentricity generated during disc formation. Eccentricity in orbital mechanics describes the ellipticity of orbits and is due to angular momentum deficit (AMD). This is the difference between the circular orbit angular momentum and  the orbital angular momentum $\mathbf{L}=\rho \mathbf{v}\times \mathbf{r}$, with  $\mathbf{v}$ the velocity of the fluid element and $\mathbf{r}$ its position relative to the sink particle, 
\begin{align}
    \mathrm{AMD}=|\mathbf{L}_\mathrm{circular}-\mathbf{L}|,
\end{align}
which is a conserved quantity in an ideal adiabatic disc.
In this study, we applied well established methods used for the calculation and analysis of disc eccentricities in past studies, such as \citet{Ragusa_2020,ragusa:24}, \citet{Teyssandier_2017}, and \citet{Lynch_2023}.
Unlike previous works our discs are formed self-consistently by collapse, this does, however, mean that the discs studied (at times $~5-20~\mathrm{kyr}$ after sink formation) are still deeply embedded. To measure the eccentricity of the disc, we must first isolate the disc. The disc isolation is done based on a  local angular momentum criterion the disc is isolated at the $|\mathbf{L}|=0.001 |\mathbf{L}|_\mathrm{max}$ isosurface, where $|\mathbf{L}|_\mathrm{max}$ is the maximum local angular momentum in the computation domain. The disc eccentricity is then calculated on this extracted disc; however, it is important to note that the quality of the disc isolation does not affect the eccentricity calculation, as eccentricity is present throughout the disc, not only in the outer disc, which may or may not be cut depending on disc selection criterion (see Figure~\ref{fig:disc_dens}).
The eccentricity is calculated using the eccentricity vector $\mathbf{e}$ (which has magnitude of eccentricity and direction pointing towards the longitude of pericentre), defined as:
\begin{align}
    \mathbf{e}=\frac{\mathbf{v}\times\mathbf{l}}{\mu}-\mathbf{\hat{r}},
\end{align}
where $\mathbf{l}$ is specific angular momentum, $\mu$ is the gravitational parameter, $\mathrm{G}M_\mathrm{star}$, and $\mathbf{\hat{r}}$ is the normalised position $\frac{\mathbf{r}}{|\mathbf{r}|}$. In eccentric systems, a ring of constant radius intersects multiple orbits, so to obtain an averaged eccentricity, we average $\mathbf{e}$ over the semi-major axis, $a$, instead, where $a$ is defined as
\begin{align}
    a=\left(\frac{2}{|\mathbf{r}|}-\frac{\mathbf{v}^2}{\mu}\right)^{-1}.
\end{align}
We used averaged eccentricity to account for pressure effects on the orbital velocity. In a disc with pressure support, there is always an angular momentum deficit, even on a circular orbit, as the orbital velocity is sub-Keplerian. This results in the eccentricity vector always having a contribution pointing towards the central object in a partly pressure supported disc \citep{Ragusa_2018}. Averaging $\mathbf{e}$ around an orbit then accounts for this effect as all the spurious centrally pointing contributions average away \citep{Teyssandier_2017}. Once the eccentricity vector is orbitally averaged the eccentricity $e$ is defined as 
\begin{align}
    e=|\langle\mathbf{e}\rangle_a|,
\end{align}
where $\langle Q \rangle_a$ is the orbital average of a quantity $Q$, 
\begin{align}
    \langle Q \rangle_a = \frac{\int_{V}w\left({a\left(\mathbf{r}\right),a}\right)Q\left({\mathbf{r}}\right)\mathrm{d}\mathbf{V}}{\int_{V}w_{a\left(\mathbf{r}\right),a}\mathrm{d}\mathbf{V}}
\end{align}
with the weight function 
\begin{align}
    w\left({a\left(\mathbf{r}\right),a}\right)=\begin{cases}
    1 & \text{if $a_\mathrm{min} < a\left(\mathbf{r}\right) \leq a_\mathrm{max}$}\\
    0 & \text{else},
    \end{cases}
\end{align}
where the semimajor axis averaging is done in bins defined by $a_\mathrm{min}$ and $a_\mathrm{max}$ to account for the finite size of computational cells.
\section{Results}
\subsection{Young disc morphology}
\begin{figure}
    \centering
    \includegraphics[scale=0.2]{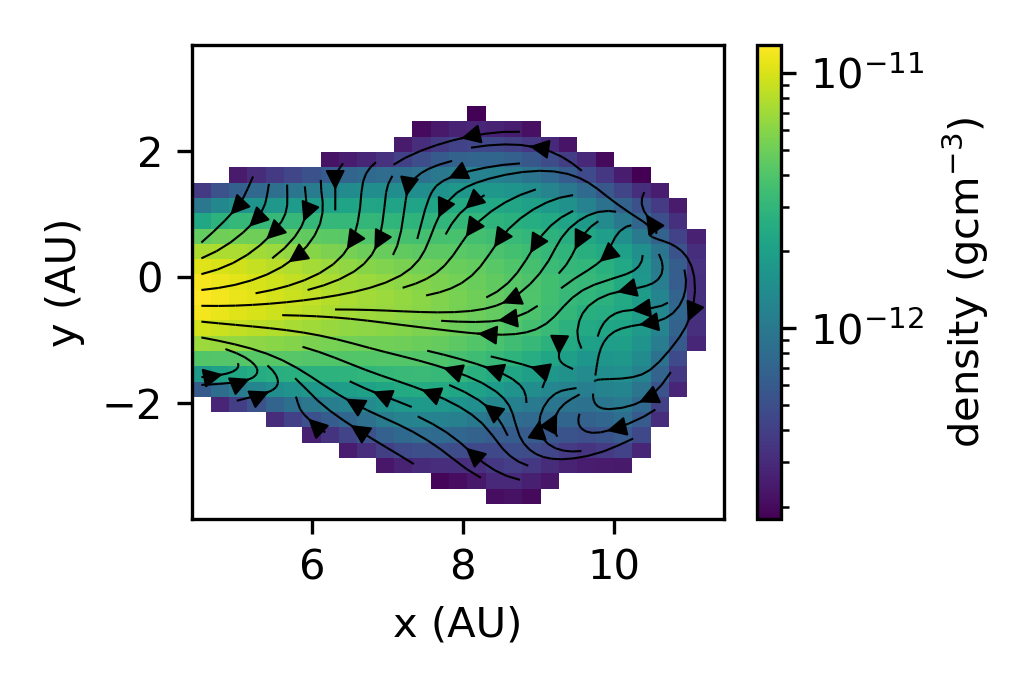}
    \caption{Plot of a vertical slice of the outer disc density of the high resolution zoom in, velocity is overplotted in streamlines, showing vertical motion in the outer disc. }
    \label{fig:vert_mix}
\end{figure}

In our simulations, we formed low-mass discs ($0.012~\mathrm{M}_\odot$) around a low mass sink of $M_\mathrm{sink}=0.26~\mathrm{M}_\odot$.
The discs formed in our simulations are characterised by a high $H/r$ ratio (see figure~\ref{fig:disc_dens} panel 1), due to heating from the protostar and compression of the infalling gas. The high disc temperature stabilises the discs against gravitational instability and the maximum Toomre Q in the disc is greater than 1. This is in contrast to other results, which found that discs  are self-regulated by GI \citep{Xu_2021}. In our case, as in \citet{hennebelle:20}, the disc is regulated by magnetic braking and ambipolar diffusion --  and not GI. The infall onto the disc is dominated by streamer infall on the disc edge and infall onto the very inner disc and protostar. The disc also exhibits strong vertical mixing (fig.~\ref{fig:vert_mix}), which may be due to baroclinic instability \citep{Charney_1947, Lesur_2010} or the parametric instability \citep{Bayly_1986, Gammie_2000, Papaloizou_2005}. This prevents dust in the discs from settling, and strong temperature gradients from arising, resulting in an evenly warm disc with a flat dust to gas ratio similar to the discs observed in the eDisk survey \citep{ohashi:23}.

\subsection{Measured eccentricity}
\begin{figure}
    \includegraphics[scale=0.1]{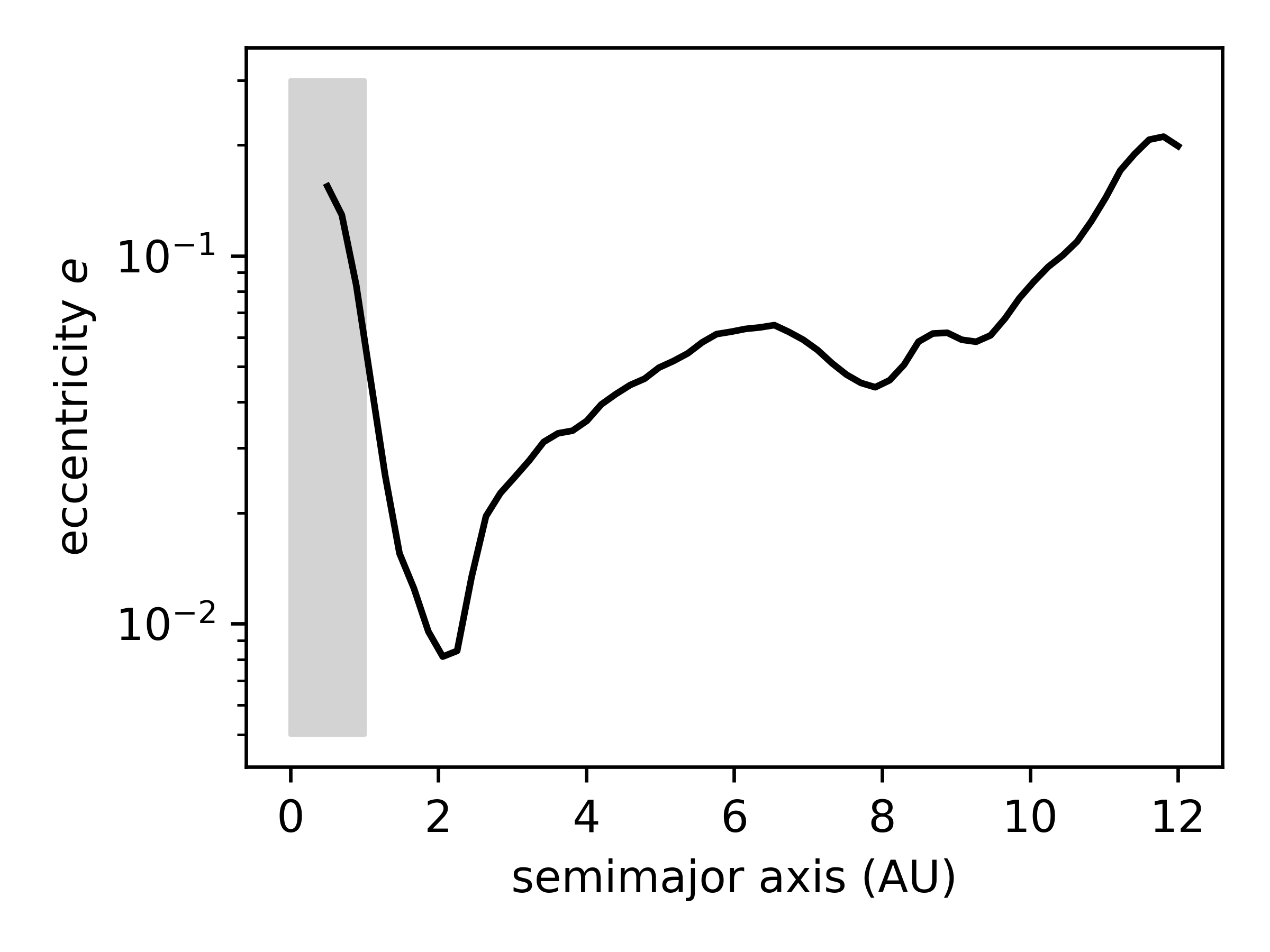}
    \caption{Eccentricity distribution 15 kiloyear after sink formation, obtained from zoom-in simulation (0.25AU resolution in the disc). The greyed-out area falls within the sink radius and should be taken with caution. The disc shows strong eccentricity throughout the disc with very high eccentricity in the outer disc, up to an eccentricity of 0.2.}
    \label{fig:disc_eccentricity_high_res}
\end{figure}

\begin{figure*}
    \centering
    \includegraphics[scale=0.1]{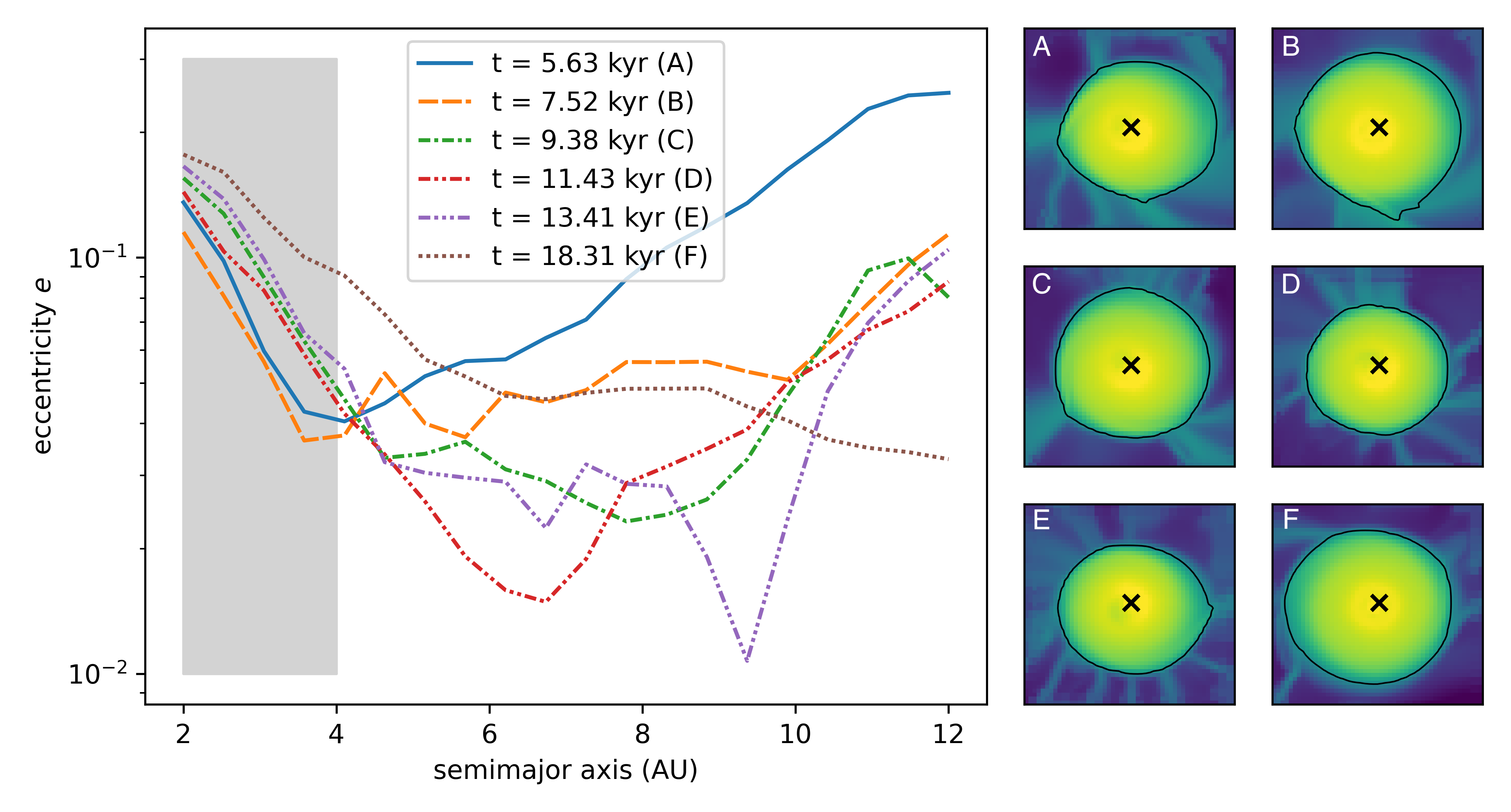}
    \caption{Plot of disc eccentricity over time in fiducial simulation. The left panel shows eccentricity profiles, the greyed out area falls within the sink radius and should be taken with caution. Right panels show the density maps of the disc (face on view) at each time where the eccentricity was calculated (from left to right and then top to bottom chronologically). On these panels, the cross marks the sink position (not to scale) and the fine black line marks the disc cut isosurface. The outer disc eccentricity decays with time and appears to evolve towards a stable eccentric disc mode.}
    \label{fig:disc_eccentricity}
\end{figure*}

The discs produced in our simulations are measurably eccentric despite the initial conditions being axisymmetric. We observed maximum eccentricities in excess of 0.2 in the high resolution zoom-in restarts (see Fig.~\ref{fig:disc_eccentricity_high_res}) and of about 0.1 in the low resolution fiducial simulation (see fig.~\ref{fig:disc_eccentricity}). On scales larger than the sink accretion radii, the shape of the eccentricity profile is similar in both cases. The eccentricity damps with time (see Fig.~\ref{fig:disc_eccentricity}) and evolves towards the fundamental eccentric disc mode. In Fig.~\ref{fig:Ap1},  we can see  the alignment of the pericenter longitudes, as expected for a disc dominated by a fundamental mode. Eccentricity damping is (at least in part) due to grid dissipation, as even in our zoom-in simulations the disc is quite coarsely resolved in comparison to work focusing on eccentricity evolution in global discs like \citet{Lynch_2023}. It must be noted that the eccentricity at $a>11$~au in the zoom-in simulations may be overestimated,  due to the way we computed the eccentricity, which does not fully resolve the eccentricity gradient. This results in some orbits not being closed, making it harder to  average away pressure effects using the method described in Sect. \ref{sec:methods:ecc}. It is worth noting that the lower resolution simulation does not exhibit this behaviour, as the sharp eccentricity gradient is not resolved and no open orbits appear. 
\begin{figure}
    \centering
    \includegraphics[scale=0.2]{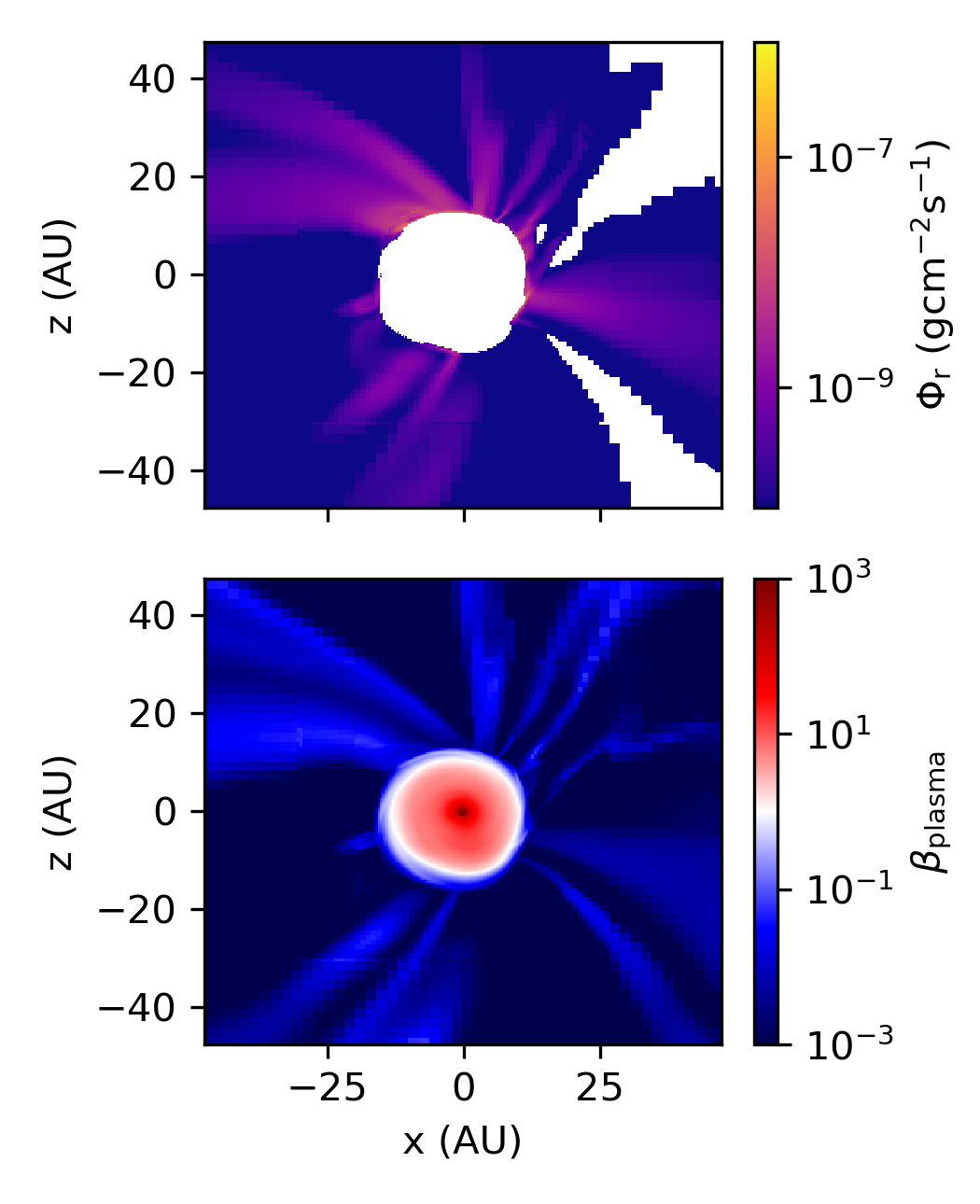}
    \caption{Radial mass flux (top) showing non-axisymmetric accretion onto the disc, which drives the development of eccentricity in the disc. Plasma $\beta$ (bottom) showing the disc midplane to be mostly magnetically dead, except the outermost regions of the disc. The accretion streams are also notably much less magnetically dominated than their surroundings.}
    \label{fig:disc_beta_phi}
\end{figure}

\section{Discussion}
\subsection{A mechanism for eccentricity excitation}
The eccentricity of the discs could be explained by several mechanisms: gravitational interactions between disc and sink \citep{Vos_2015}, an eccentricity pumping fluid instability like the viscous overstability \citep{Kato_1978,Papaloizou_1986,Latter_2006}, eccentric gravitational instability \citep{adams:89,lin:2015,li:21}, or non-axisymmetric accretion as has been seen in simulations of tidal disruption events \citep{Guillochon_2014,Shiokawa_2015,Bonnerot_2020}. In this study, we find strong evidence for the last mechanism. The infalling material self arranges into non-axisymmetric streams, which feed the disc with high eccentricity material, thereby increasing the disc eccentricity. This mode of stream accretion can be seen in Fig.~\ref{fig:disc_beta_phi}, with  the top panel showing the mass flux onto the disc edges being concentrated into streams. The bottom panel in Fig.~\ref{fig:disc_beta_phi} shows $\beta_\mathrm{plasma}$, defined as $\beta_\mathrm{plasma}=P_\mathrm{thermal}/P_\mathrm{magnetic}$. The $\beta_\mathrm{plasma}$ is much higher in the streams than the surrounding material, highlighting the non-axisymmetric structure. This also suggests that MHD effects may play a role in funnelling material into the streams. It is these accretion streams that deliver AMD to the disc. The infalling material must have high AMD as it is arriving on what is basically a parabolic trajectory, if the infall was axisymmetric the AMD would sum to 0 around the disc and no eccentricity would be excited. In our simulations, however, the infall is localised in streams, resulting in the discs' AMD increasing and eccentricity being excited as a result. We note that the AMD can be lost or gained from open inner boundary; whereas, is conserved for  reflective boundaries, even when slowly accreting  \cite{kley:08}. In our case, the effect of the sink accretion scheme on the conservation of AMD remains to be investigated in future works.
\begin{figure}
    \centering
    \includegraphics[scale=0.2]{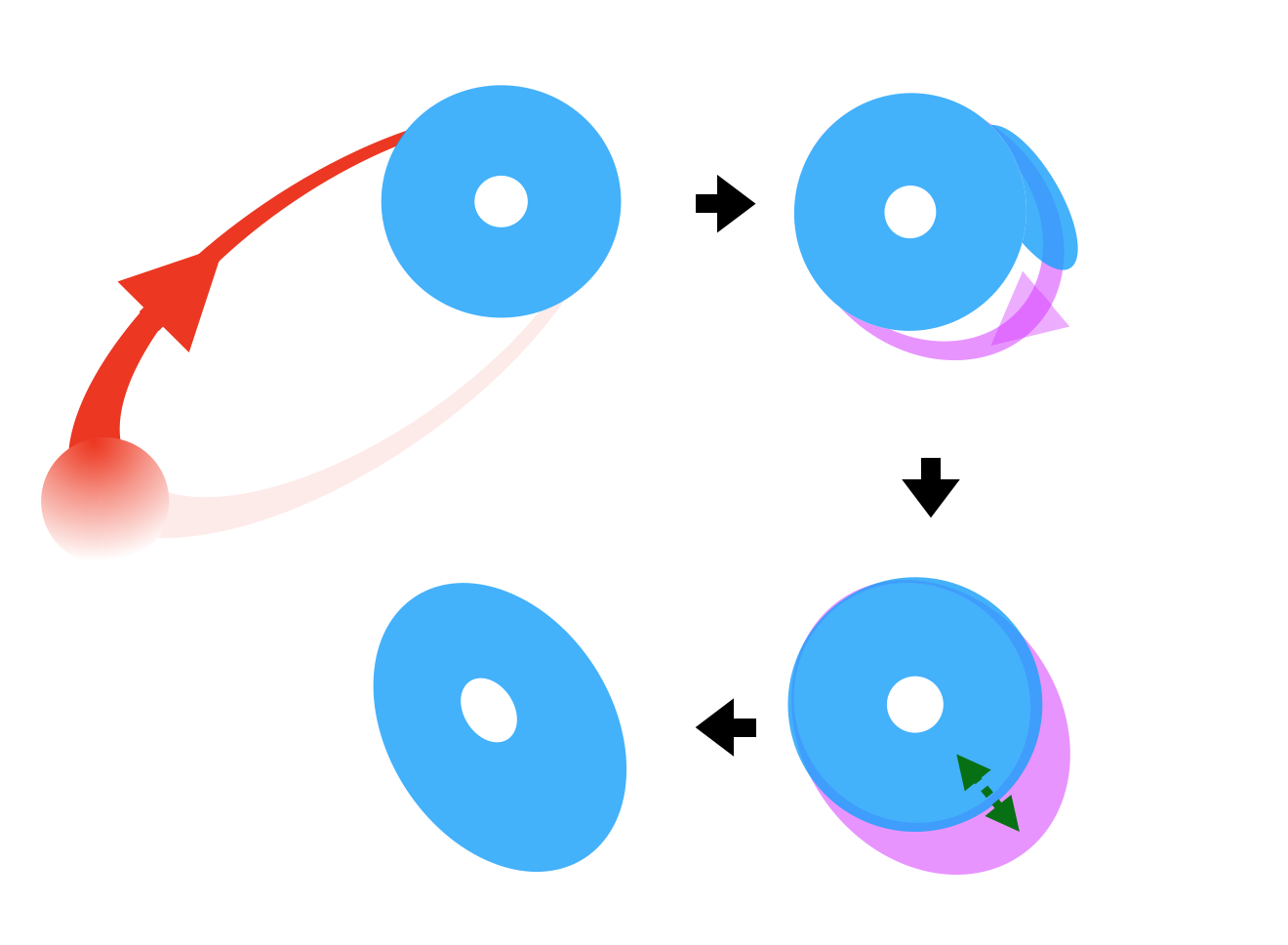}
    \caption{Sketch of the mechanism of eccentricity excitation due to non-axisymmetric infall of material on parabolic orbits. Following the arrows: 1. Material infalls onto the disc on an effectively very eccentric orbit. 2. This infalling material carries large AMD, so when acreted it increases the AMD in the discand excites material in the outer orbits of the disc. 3. The resulting disc has an eccentric outer disc and more circular inner disc. 4. Through AMD transport the disc settles into a long-lived configuration.}
    \label{fig:sketch}
\end{figure}

To explain the results of this study, we need a mechanism responsible for breaking axisymmetry. A likely culprit is self-gravity, but it appears that MHD plays a role in the symmetry breaking as well. This can be inferred from the $\beta_\mathrm{plasma}$ (Fig.~\ref{fig:disc_beta_phi} bottom panel), where the accretion streams have $\beta_\mathrm{plasma}~10^{-1}$, while in the surroundings, we have  $\beta_\mathrm{plasma}<10^{-3}$. The gas therefore drags the magnetic fields in the streamers leading to a funnelling of the accreting material in the streamers. The funnelling is also further enhanced by the self gravity of the streamer. The final result being that accreting material organises itself into thin streamers, which guarantee non-axisymmetric delivery of mass on highly eccentric trajectories. This enhances the total AMD in the disc and drives the eccentricity. This filamentation of streamers has been also observed in previous numerical works \citep[e.g.][]{seifried:15,mignon:21a,tu:24}, as well as in observations \citep{flores:23,kido:23}. For a review, we refer to \cite{pineda:23}. In particular, \cite{kido:23} reported  high-resolution eDisk observations of the Class 0 protostar IRAS 16544. They identified a compact Keplerian disk ($< 30$ au), which exhibits flared and non-naxisymmetric dust distribution and which is associated to non-axisymettric  accretion streamers. Altogether, their results are very consistent with the scenario we propose in this letter. 
\subsection{Disc eccentricities: Implications}
The results found in this study suggest that young discs should have detectable eccentricity of $e_\mathrm{max}>0.1$. This eccentricity may be hard to detect in Class 0 discs due to the deep embedding, but the eccentric modes should remain trapped in the disc \citep{Lee_2019}, meaning that the eccentricity should damp on the viscous timescale of the disc.  Thus, even in turbulent discs with $\alpha_\mathrm{bulk} \sim 10^{-3}$, this should result in the eccentricity surviving into Class I and should become observable. It is worth noting that while the AMD is conserved, it is transported within the disc potentially leading to eccentricity profiles that are harder to detect \citep{huang:18},  particularly given the larger disk sizes in Class II \citep[e.g.][]{najita:18}. In our simulations, eccentricity does indeed decay over time (see Fig.~\ref{fig:disc_eccentricity}), but this is known to result from grid level dissipation and dependant on resolution \citep{Lynch_2023,Teyssandier_2019}, which is supported by the higher eccentricity seen in the zoom in simulation. This result suggests that we need a 'somewhat' efficient eccentricity dissipation mechanism to explain the very circular observed Class II and late Class I  discs \citep{andrews:18}. The problem is that strong bulk viscosity is required for the disc to dissipate AMD efficiently and there is no other obvious way to damp AMD. The parametric instability may drive sufficient turbulence to damp eccentricity \citep[see][]{Wienkers_2018}. This theory makes a strong prediction on dust settling and turbulent broadening in early Class I discs, in that the dust cannot be well settled and the turbulent broadening must be consistent with strong turbulence $\alpha_\mathrm{bulk}>10^{-3}$. This also means that the dust settling must be fast after this turbulent circularisation epoch ends to match constraints from Class II discs \citep{Rosotti_2023}. 
\section{Conclusions}
Our simulations show young discs forming with eccentricities of above 0.1. A combination of MHD effects and self gravity break axial symmetry in the collapse. The resulting non-axisymmetric accretion leads to excitation of eccentricity in the disc. Eccentricity is not efficiently dissipated in the disc, despite grid effects in our Cartesian grid simulations. The outer disc is observably turbulent, which could be due to either baroclynic or parametric instabilities as the disc meets the instability criteria for both. The result of this turbulence is that the discs are well mixed and are not stratified. The discs we observe appear quite similar to the eDisk survey observations, in that the discs are compact and vertically well mixed. We find that the observations of, for example, \cite{kido:23} of non-axisymmetric discs are in good agreement with our simulations.
In summary, this study makes several predictions for moderate mass protostellar discs:
\begin{itemize}
    \item Young discs should have eccentricities of about 0.1 or higher and as a result should have moderate to strong axial asymmetry.
    \item Young discs should be turbulent and lack vertical stratification.
\end{itemize}
\begin{acknowledgements}
FL and BC were supported by the French national research agency grant ANR-20-CE49-0006 (DISKBUILD), EL and ER were supported by ERC No. 864965 (PODCAST) under the EU’s Horizon 2020 programme, ER was also supported by the Marie Sk\l{}odowska-Curie grant No. 101102964 (ORBIT-D) under the EU's Horizon Europe programme. We acknowledge Eline Maaike De Weerd for her contribution through extensive proofreading. We gratefully acknowledge support from the PSMN (P\^{o}le Scientifique de Mod\'{e}lisation Num\'{e}rique) of the ENS de Lyon for the computing resources. This work utilised resources from DARI through allocations A0110407247 and AD010414051. {\bf Author contributions: } FL lead the entire work, from simulation design, run and postprocessing to the manuscript writing. BC, EL and ER contributed to the simulation postprocessing, figure production and manuscript drafting.
\end{acknowledgements}

\bibliographystyle{aa}
\bibliography{biblio}

\appendix
\section{Pericentre alignment \label{sec:appendixA}}
\begin{figure}
    \centering
    \includegraphics[scale=0.07]{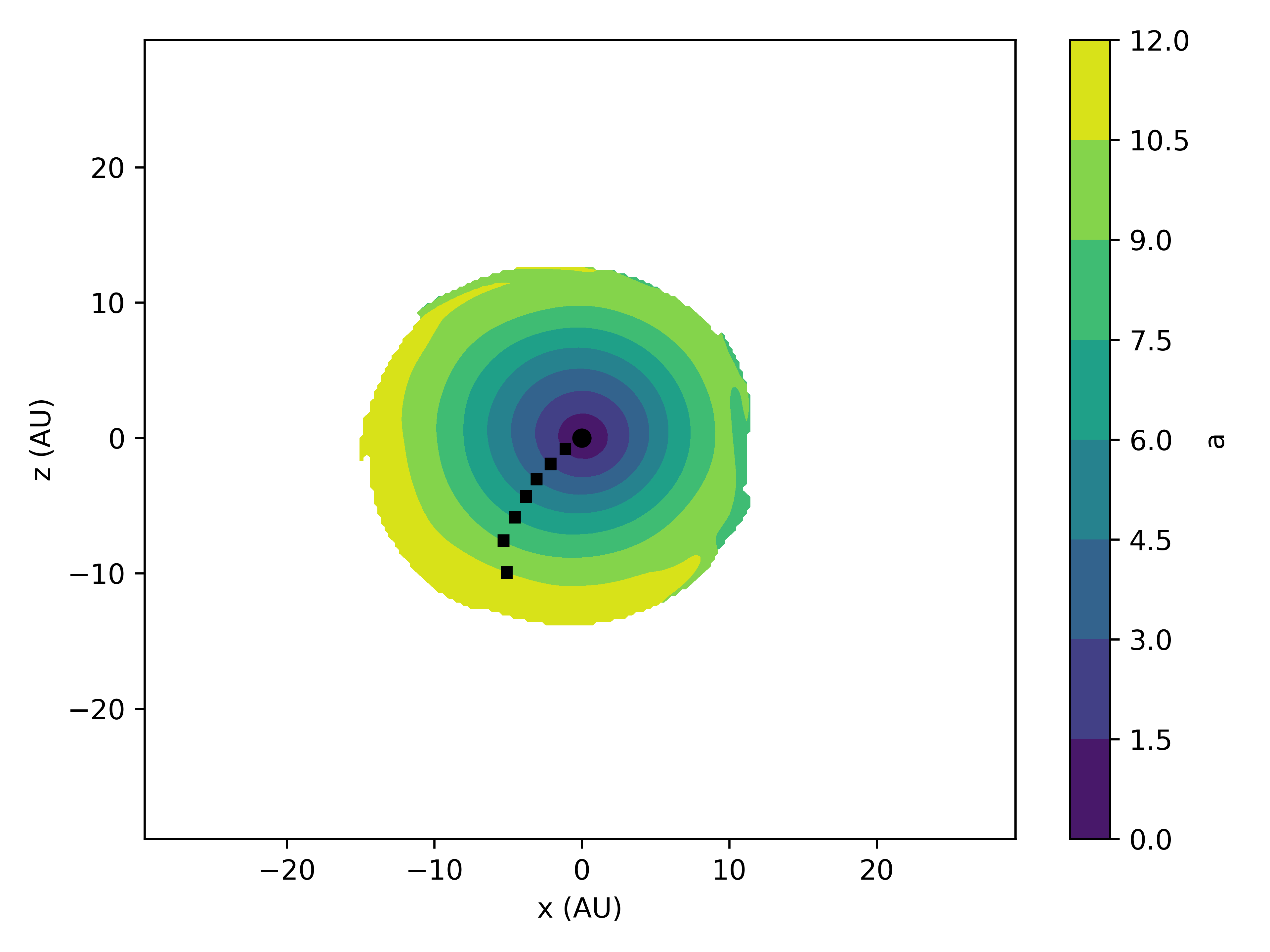}
    \caption{Pericentre locations at t = 15 kyr in the zoom-in simulation. The pericentres are aligned. A crescent arises where material accretes onto the disc  }
\label{fig:Ap1}
\end{figure}

\begin{figure}
    \centering
    \includegraphics[scale=0.07]{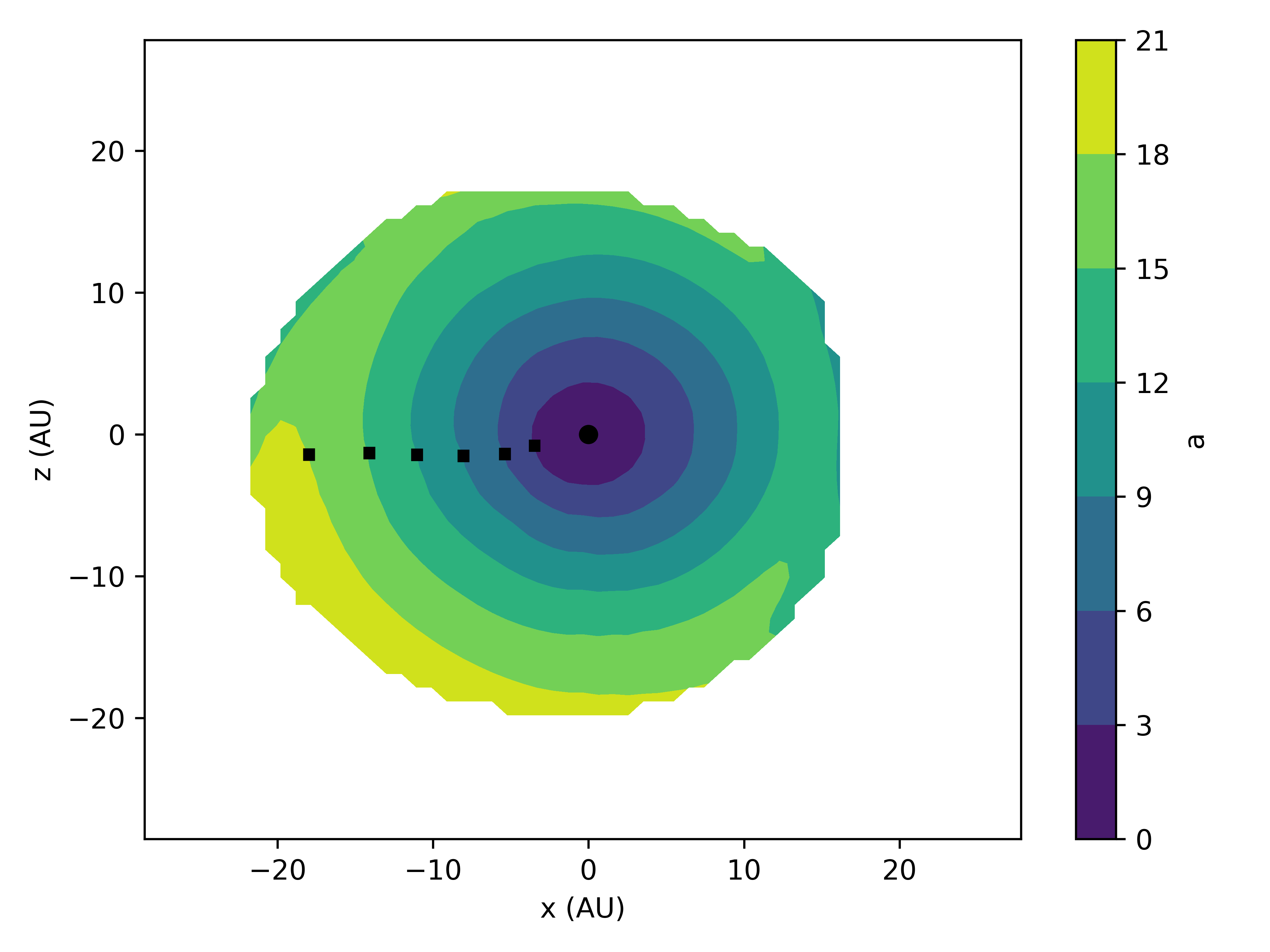}
    \caption{Pericentre locations at t = 18 kyr in the fiducial case. The pericentres are well aligned and and a crescent can be seen at the location of a large inflow.}
\label{fig:Ap2}
\end{figure}

To determine whether the discs are settling towards an eccentric mode, we show  (in Fig. \ref{fig:Ap1} and \ref{fig:Ap2}) that the pericentres of the ellipses, defined as the point minimising distance from the centre of mass for a given ellipse, align. They maintain this alignment as the disc precesses. This behaviour is only exhibited when the disc is dominated by the fundamental eccentric mode \citep{goodchild:06}, which is consistent with the shape of the late time eccentricity profiles in Fig. \ref{fig:disc_eccentricity}. Non-modal eccentric discs would be expected to exhibit differential precession, while discs dominated by higher order modes would have anti-aligned pericentres over part of the disc.

\end{document}